\documentclass{cjpsuppl}% for LaTeX 2e
\usepackage{pstricks}
\usepackage{graphpap}
\usepackage{graphicx}% Include figure files
\newcommand{\sherpa}{{\tt SHERPA}}
\begin{document}
\title{Predictions for multi-particle final states with SHERPA}
\authori{T.~Gleisberg, S.~H{\"o}che, F.~Krauss, A.~Sch{\"a}licke, S.~Schumann\footnote{Talk given by S.~Schumann at Physics at LHC, 13--17 July, 2004, Vienna, Austria.}, G.~Soff, J.~Winter} \addressi{Institut f\"ur theoretische Physik, TU Dresden, D-01062 Dresden, Germany\\ E-mail: steffen@theory.phy.tu-dresden.de}
\authorii{}  \addressii{}
\authoriii{}    \addressiii{}
\authoriv{} \addressiv{}
\authorv{}       \addressv{}
\authorvi{}    \addressvi{}
\headtitle{Predictions for multi-particle final states with SHERPA}
\headauthor{T. Gleisberg et al.}

\lastevenhead{T. Gleisberg et al.: Predictions for multi-particle final states with SHERPA}
%%%%%%%%%%%%%% Pro editory supplementu: %%%%%%%%%%%%%%%
\refnum{}%slouzi editorum pro evidenci; nakonec {}
\daterec{10 September 2004}
\suppl{??}  \year{2222} \setcounter{page}{1}
%\firstpage{1}
%\lastpage{000}
%\makefirsttitle
%%%%%%%%%%%%%%%%%%%%%%%%%%%%%%%%%%%%%%%%%%%%%%
\maketitle

\begin{abstract}
In this contribution the new event generation framework \sherpa\ will 
be presented, which aims at a full simulation of events at current and 
future high-energy experiments. Some first results related to the production 
of weak vector bosons in association with jets at the Tevatron and the LHC will be 
discussed.
\end{abstract}

\section{Introduction}
Multi-particle and multi-jet final states will be of enormous importance at the LHC. 
They serve both as signals for interesting physics and as important backgrounds 
to many new physics channels. As an example for the first case, the production and 
decay of top quarks or SUSY particles can be mentioned. A typical background is 
the production of weak vector bosons in association with jets. 
The simulation of such processes is a great challenge that requires the development 
of new appropriate tools, for a recent review of these developments see 
\cite{Mangano:2003ps}. 
The multi-purpose event generator \sherpa\ \cite{Gleisberg:2003xi} is one of these 
new tools. As the re-writes of the well-established tools Pythia 
\cite{Sjostrand:2000wi} and Herwig \cite{Corcella:2000bw_and_2002jc}, namely Pythia7 
\cite{Bertini:2000uh} and Herwig++ \cite{Gieseke:2004ag_Gieseke:2003hm}, 
it is entirely written in the object-oriented programming language C++. 
One of the striking features of \sherpa\ is the inclusion of the CKKW prescription 
to combine tree-level multi-jet matrix elements with parton showers 
\cite{Catani:2001cc_Krauss:2002up}. This method allows for a consistent description 
of multi-jet final states and a combination of such higher order calculations with 
the non-perturbative regime of hadron production in an universal manner. 
In Sec.~\ref{sherpa_intro} the generator \sherpa\ will be briefly reviewed. 
In Sec.~\ref{sherpa_results} results obtained for the production of weak gauge bosons 
in association with jets at the Tevatron and the LHC will be presented.
\section{The \sherpa\ generator\label{sherpa_intro}}

\begin{table}[ht]
\begin{center}
{\small
\begin{tabular}{|c|c|c|c|c|c|c|c|c|}
\hline
\multicolumn{2}{|c|}{ X-sects (pb)} & \multicolumn{6}{c|}{ Number of jets}\\\hline
\multicolumn{2}{|c|}{ $ e^- \bar \nu_e $ + $n$ QCD jets }& 0 & 1 & 2 & 3 & 4  & 5 \\
\hline
\multicolumn{2}{|c|}{Alpgen\cite{Mangano:2002ea}}   & 3904(6)& 1013(2) & 364(2)& 136(1) & 53.6(6) & 21.6(2) \\
\multicolumn{2}{|c|}{CompHEP\cite{Pukhov:1999gg}}  & 3947.4(3)& 1022.4(5)& 364.4(4)& & & \\
\multicolumn{2}{|c|}{GR@PPA\cite{GRAPPA}}   & 3905(5)& 1013(1)& 361.0(7)& 133.8(3) & 53.8(1) & \\
\multicolumn{2}{|c|}{MadEvent\cite{Stelzer:1994ta}} & 3902(5)& 1012(2)& 361(1)& 135.5(3) & 53.6(2) & \\
\multicolumn{2}{|c|}{Sherpa}   & 3908(3) & 1011(2) & 362(1) & 137.5(5) & 54(1)  &\\
\hline
\end{tabular}\\[5mm]
\begin{tabular}{|c|c|c|c|c|c|c|c|}
\hline
\multicolumn{2}{|c|}{ X-sects (pb)} & \multicolumn{5}{c|}{ Number of jets}\\\hline
\multicolumn{2}{|c|}{ $ e^- \bar \nu_e $ + $b\bar b$ }& 0 & 1 & 2 & 3 & 4  \\
\hline
\multicolumn{2}{|c|}{Alpgen}   & 9.34(4)& 9.85(6)& 6.82(6)& 4.18(7)& 2.39(5) \\
\multicolumn{2}{|c|}{CompHEP}  & 9.415(5)& 9.91(2)& & &  \\
\multicolumn{2}{|c|}{MadEvent} & 9.32(3)& 9.74(1)& 6.80(2)&  & \\
\multicolumn{2}{|c|}{Sherpa}   & 9.37(1) & 9.86(2) & 6.87(5) &   &   \\
\hline
\end{tabular}
}
\caption{\label{mc4lhc_results}Compilation of results for cross sections of some selected 
processes at the LHC. For details of the calculational setup and more results, cf.\ 
the MC4LHC homepage \cite{MC4LHC}.}
\end{center}
\end{table}

In its current version \sherpa\ is able to simulate electron--positron annihilations, 
unresolved interactions of photons with photons or leptons, 
and fully hadronic, i.e. proton--antiproton and proton--proton, collisions. 
Within \sherpa\ the different phases of the event simulation are hosted by 
physics-oriented modules. In its current version, {\tt SHERPA-1.0.4} 
\cite{sherpa_homepage}, the following physics modules are implemented:
\begin{itemize}
\item Interface to various PDFs: CTEQ \cite{Pumplin:2002vw} and MRST \cite{Martin:1999ww} 
      in their original form as well as many other PDFs through LHAPDF in its version 1 
      \cite{LHAPDFv1}.
\item {\tt AMEGIC++} \cite{Krauss:2001iv} as generator for the matrix elements for hard 
      scattering processes and decays as well as an internal library of analytical 
      expressions for some very constrained set of $2\to 2$ processes. Besides the full SM, 
      {\tt AMEGIC++} contains the full MSSM and an ADD model of large extra dimensions 
      \cite{Gleisberg:2003ue}. The SUSY particle spectra are provided by an interface to 
      Isajet \cite{Baer:1999sp}. The next \sherpa\ release will in addition support the 
      SUSY Les Houches accord interface \cite{Skands:2003cj}. 
      {\tt AMEGIC++} has exhaustively been tested for a large number of production cross 
      sections for six-body final states at an $e^+e^-$ collider 
      \cite{Gleisberg:2003bi} and for various processes at the LHC, 
      see Tab.~\ref{mc4lhc_results} and \cite{MC4LHC}.
\item For multiple QCD bremsstrahlung, i.e.\ the emission of secondary partons, 
      \sherpa's own parton shower module {\tt APACIC++} \cite{Kuhn:2000dk} is 
      invoked\footnote{In addition to the published version, it has 
      been supplemented by parton showers in the initial state, enabling \sherpa\ to also 
      simulate events with hadronic initial states \cite{apacic-2-0}.}. 
      The merging of the hard matrix elements for multi-jet production and the subsequent 
      parton showers is achieved according to the merging procedure proposed in 
      \cite{Catani:2001cc_Krauss:2002up}, heavy quarks are treated with appropriate Sudakov 
      form factors \cite{Krauss:2003cr}.
\item Multiple parton interactions, giving rise to the ``hard'' underlying  event, 
      are currently being implemented. The corresponding module will be part of 
      the next release of \sherpa.
\item Hadronisation of the resulting partons and subsequent hadron decays so far are realized 
      by an interface to the corresponding Pythia routines. However, a new version of cluster 
      fragmentation \cite{Winter:2003tt} is ready to be fully implemented in the near future.
\end{itemize}

\section{Results for \mbox{\boldmath $W$/$Z$}+jets production\label{sherpa_results}}

The production of electroweak gauge bosons, e.g.\ $W^\pm$ and $Z$ bosons, is one of the most 
prominent processes at hadron colliders. Through their leptonic decays, they leave a 
clean signature, namely either one charged lepton accompanied by missing energy for $W$ bosons 
or two oppositely charged leptons for the $Z$ bosons. The large statistics of these processes 
at the LHC will even allow the use of these channels as luminosity monitors. In addition these 
channels occur as important backgrounds to many signal processes. Due to the importance of 
these channels they were used to validate the merging procedure implemented in \sherpa\ and to 
compare the results with those of other approaches \cite{Krauss:2004??}. 

When merging matrix elements and parton showers according to the CKKW prescription the phase 
space for parton emission is divided into the hard region of jet production accounted for by 
suitable tree-level matrix elements and the softer region of jet evolution covered by the 
parton showers. Then, extra weights are applied on the former and vetoes on the latter, 
such that the overall dependence on the separation cut is minimal. The weight attached to the 
matrix elements takes into account the terms that would appear in a corresponding parton 
shower evolution. Therefore, a ``shower history'' is reconstructed 
by clustering the initial and final state particles stemming from the tree-level matrix 
element according to a $k_\perp$-formalism \cite{Catani:1992zp}. This procedure yields nodal values, 
namely the different $k_\perp$-measures $Q^2$ where two jets have been merged into one. 
These nodal values can be interpreted as the relative transverse momentum describing the 
jet production or the parton splitting. The first ingredients of the ME weight are the strong 
coupling constants evaluated at the respective nodal values of the various parton splittings, 
divided by the value of the strong coupling constants as used in the evaluation of the 
matrix element. The other part of the correction weight is provided by suitable Sudakov form 
factors. 

\begin{figure}[h]
\begin{center}
\begin{picture}(350,245)
%\graphpaper(0,0)(350,245)
\put(223,114){\includegraphics[width=5cm]{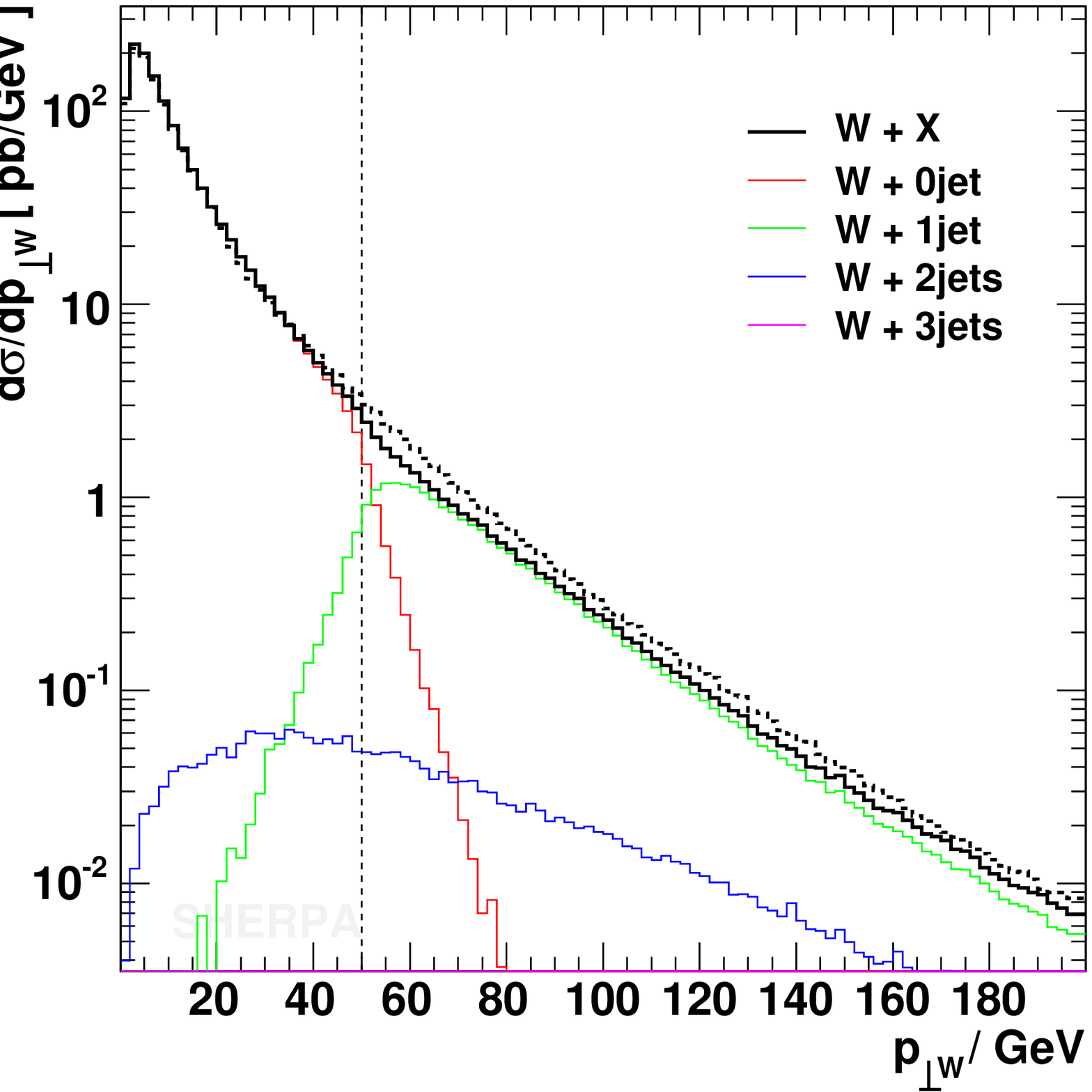}}
\put(109,114){\includegraphics[width=5cm]{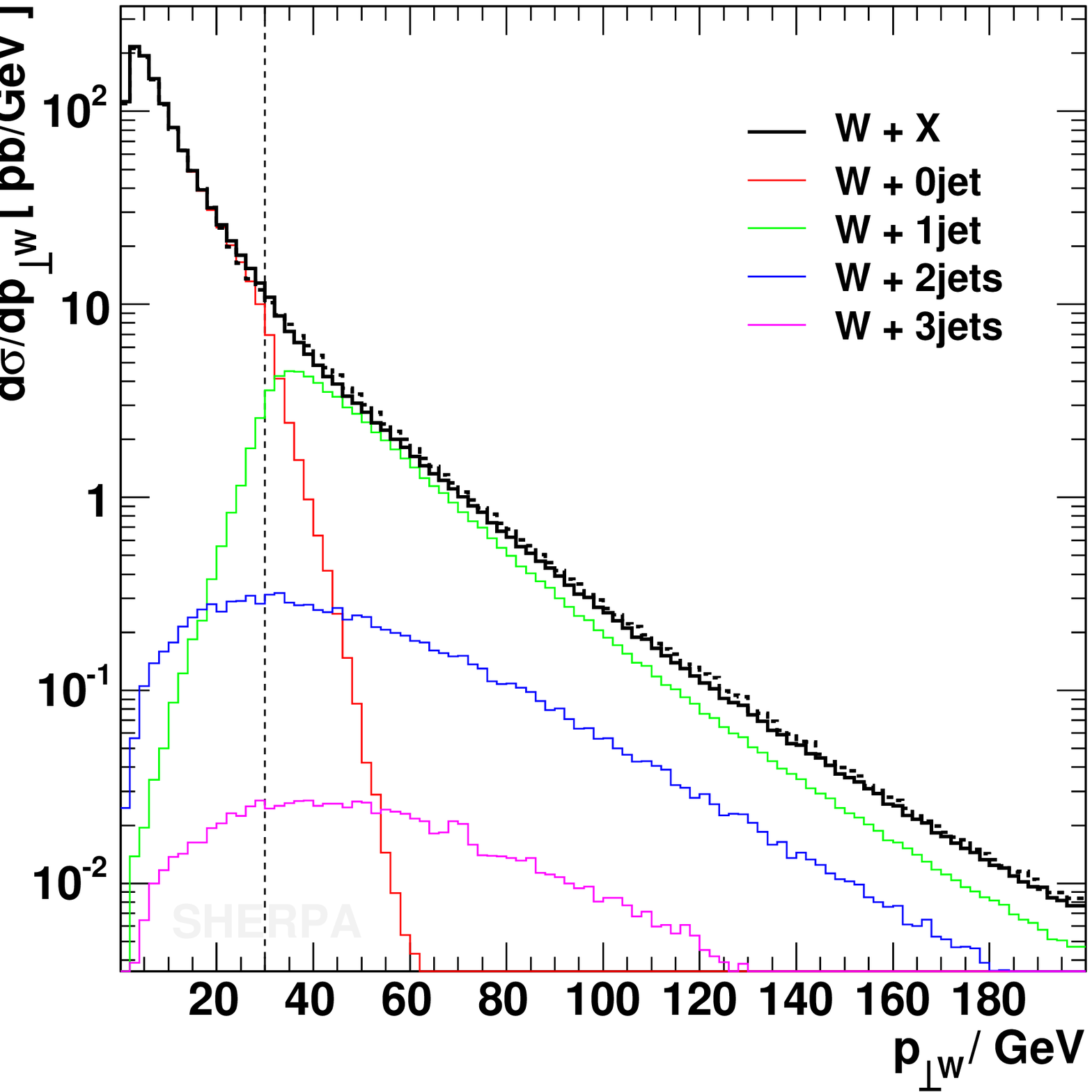}}
\put(-5,114){\includegraphics[width=5cm]{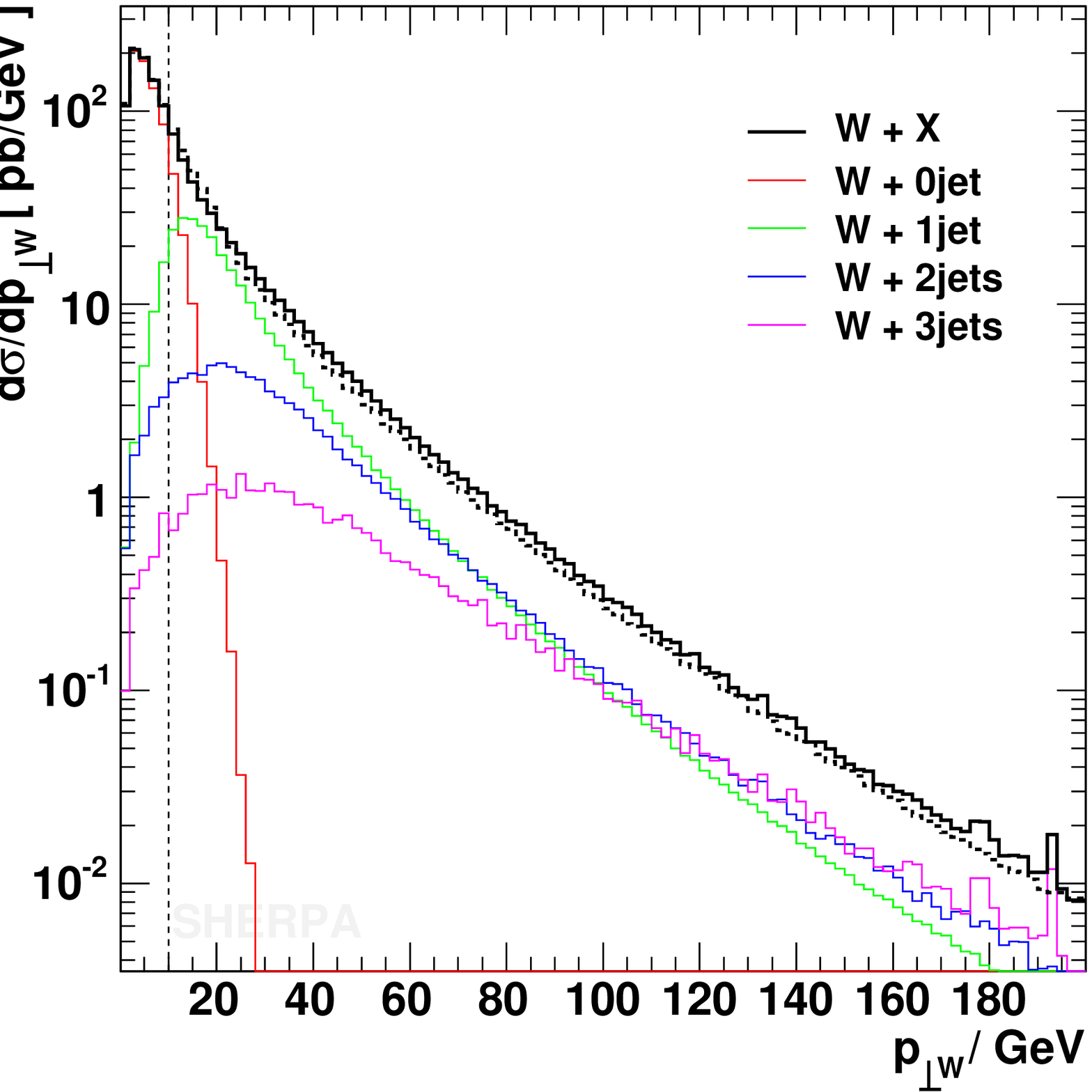}}
\put(223,0){\includegraphics[width=5.0cm]{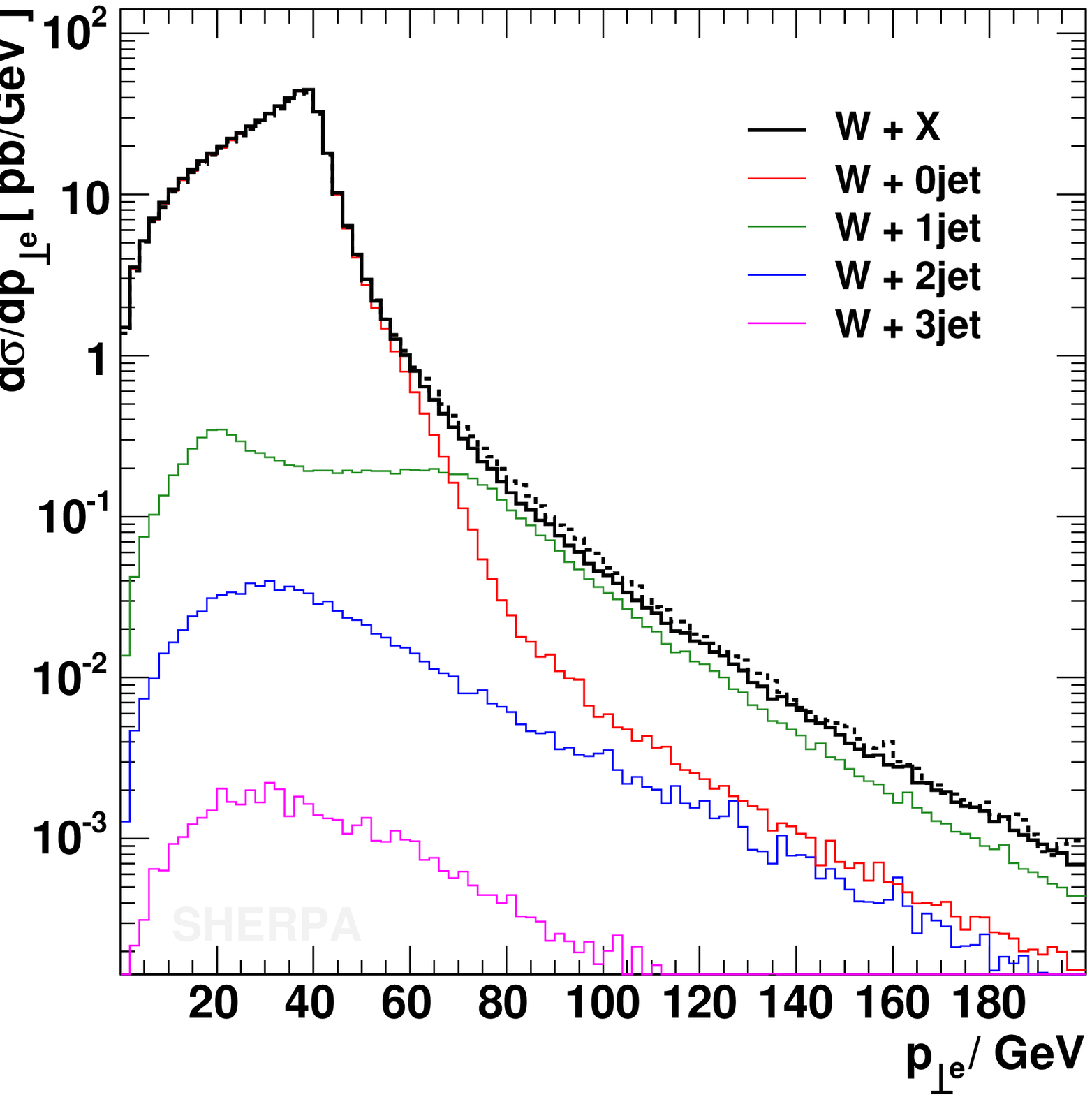}}
\put(109,0){\includegraphics[width=5.0cm]{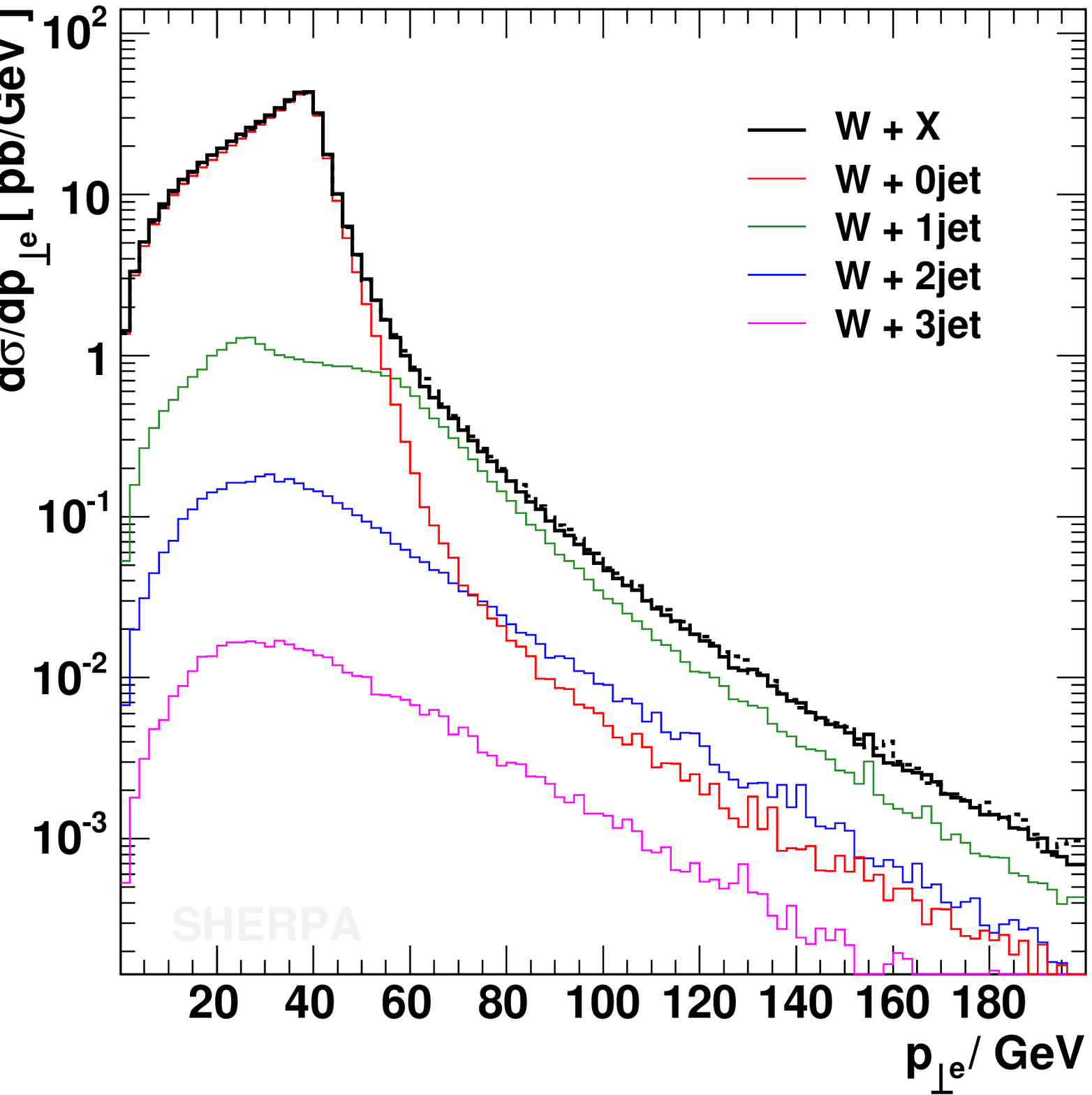}}
\put(-5,0){\includegraphics[width=5.0cm]{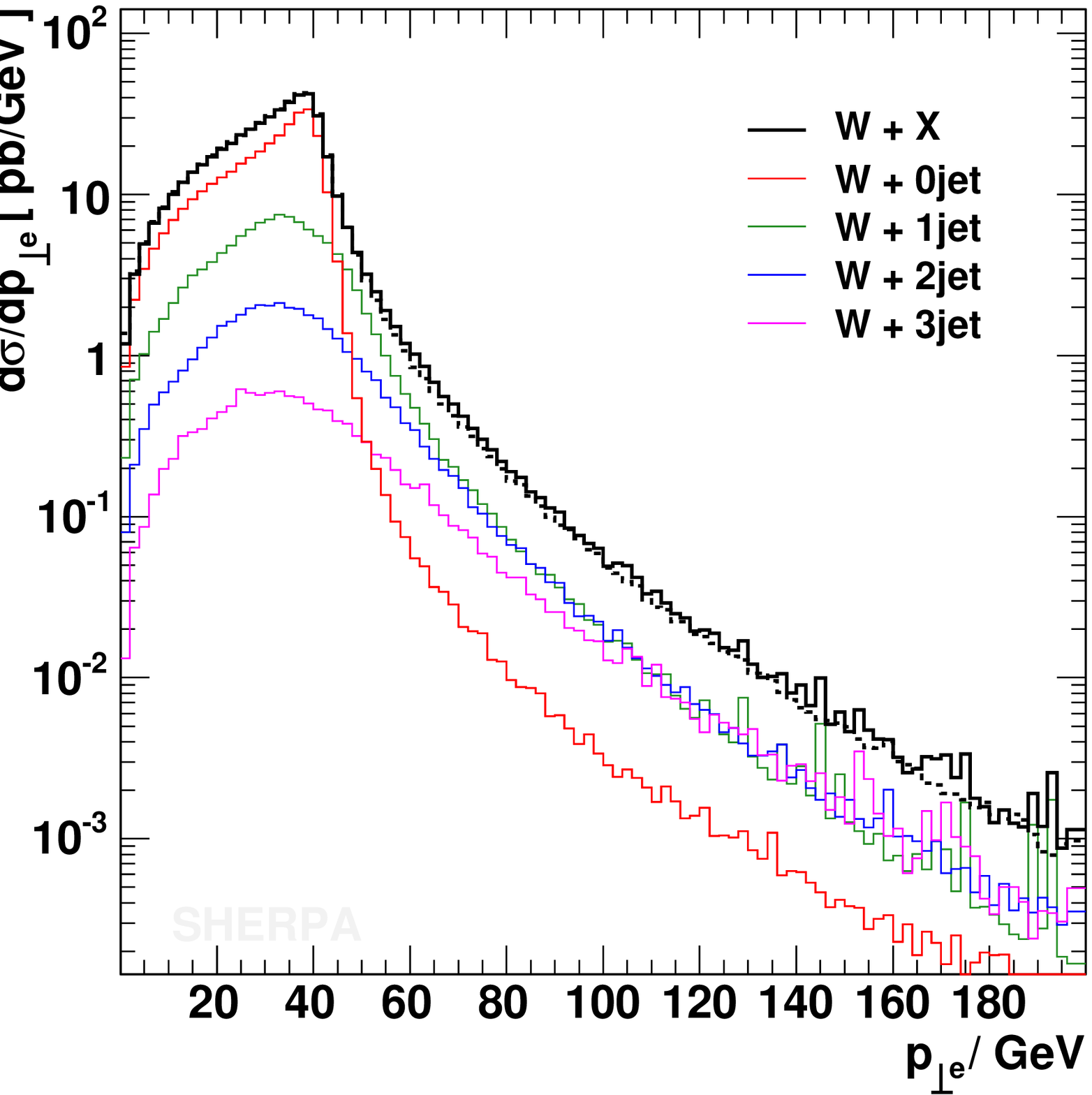}}
\end{picture}
\end{center}
\caption{\label{ycut_pt}$p_\perp(W^-)$ and $p_\perp(e^-)$ for 
         $Q_{\rm cut}=10$ GeV, $30$ GeV and $50$ GeV (black solid line) 
         in comparison with $Q_{\rm cut}=20$ GeV (black dashed line). 
         The colored lines indicate the contributions 
         of the different multiplicity processes.}
\end{figure}

The first concern that has to be proven is that the dependence of the predictions on the 
merging scale $Q_{\rm cut}$ is small. Fig.\ \ref{ycut_pt} shows the transverse momentum 
distribution of the $W^-$ boson and the corresponding electron in inclusive production at 
Tevatron Run~II. The black solid line represents the total inclusive result as obtained 
with \sherpa. A vertical dashed line indicates the respective separation
cut $Q_{\rm cut}$, which has been varied between 10~GeV and 50~GeV. 
To guide the eye, all plots also show the same observable as obtained with a 
separation cut $Q_{\rm cut}=20$~GeV, which is shown as a dashed black curve. 
The coloured lines give the contributions of different multiplicity processes. 
For the transverse momentum of the $W$ below the cut, the distribution 
is dominated by the LO matrix element with no extra jet, i.e.\ the transverse momentum 
is generated by the initial state parton shower only. Around the cut, a small dip is 
visible. The $p_\perp$ distribution of the electron, in contrast, is hardly altered.
When looking on the rapidity distribution of the $W$ boson and the electron, again 
to a very good approximation an independence on the merging scale has been observed.

In order to study the impact of the merging prescription, the predictions obtained with 
\sherpa\ have to be compared with other approaches. For $W$ and $Z$ plus jet 
production the results can be confronted with NLO QCD calculations delivered by the 
parton level generator MCFM \cite{Campbell:2002tg}. In order to compare both approaches, 
a two-step procedure is chosen. In a first step the Sudakov and $\alpha_s$ reweighted matrix 
elements are compared with exclusive NLO results obtained with MCFM. In the case of the 
next-to-leading order calculation, the exclusiveness of the final states boils down to a 
constraint on the phase space for the real parton emission. 
The exclusive \sherpa\ results consist of appropriate leading order matrix elements with 
scales set according to the $k_\perp$-clustering algorithm. The exclusiveness is achieved 
by suitable Sudakov form factors. In a second step, the jet spectra for inclusive production 
processes are compared. For the next-to-leading order calculation, this time the phase 
space for real parton emission is not restricted and the \sherpa\ predictions are obtained 
from a fully inclusive sample, using matrix elements with up to two extra jets and the 
parton showers attached. For the jet definition the Run~II $k_\perp$-clustering algorithm 
defined in \cite{Blazey:2000qt} is used.

\begin{figure}
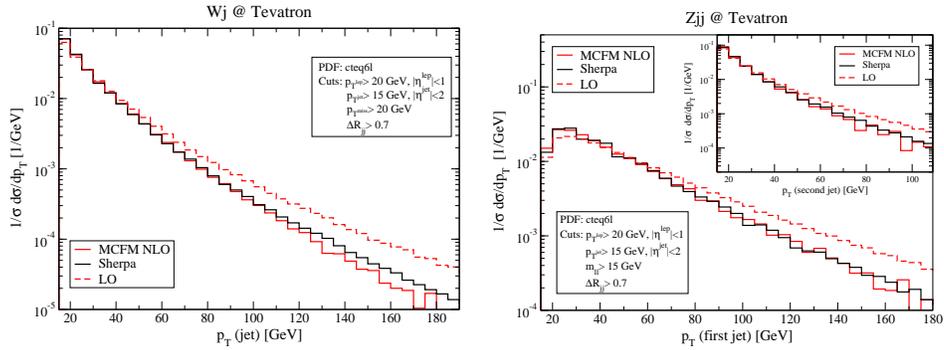

\begin{center}
\begin{tabular}{cc}
\includegraphics[width=6cm]{figures/TeV_W1jet_excl_0.7.eps}&
\includegraphics[width=6cm]{figures/TeV_Z2jet_excl_0.7.eps}
\end{tabular}
\end{center}
\caption{\label{TEV_exclusive} The $p_T$ distribution of the jet in exclusive $W$+1jet 
events (left) and the first and second jet in $Z$+2jet events (right) at 
Tevatron Run~II. The \sherpa\ predictions are compared to the naive LO results and 
the corresponding NLO QCD predictions obtained with MCFM \cite{Campbell:2002tg}.}
\end{figure}

In Fig.\ \ref{TEV_exclusive} the jet $p_\perp$ distribution for the exclusive production of 
$W$+1jet and $Z$+2jets at Tevatron Run~II are shown. In both figures, the \sherpa\ prediction 
is compared with the exclusive NLO result obtained with MCFM and with the naive LO prediction. 
For the fixed-order NLO and LO result, the renormalisation and factorisation scales have been 
set to $\mu_R=\mu_F=M_W$. All distributions have been normalised to the 
corresponding total cross section. This allows for a direct comparison of the distribution's 
shape. The change between the naive leading order and the next-to-leading order distribution 
is significant. At next-to-leading order the distributions become much softer. 
For a high-$p_\perp$ jet it is much more likely to emit a parton that fulfils the jet 
criteria and, therefore, the event is removed from the exclusive sample. The \sherpa\ 
predictions show the same feature. The inclusion of Sudakov form factors and the scale 
setting according to the merging prescription improves the LO prediction, leading to a 
rather good agreement with the next-to-leading order result.

\begin{figure}
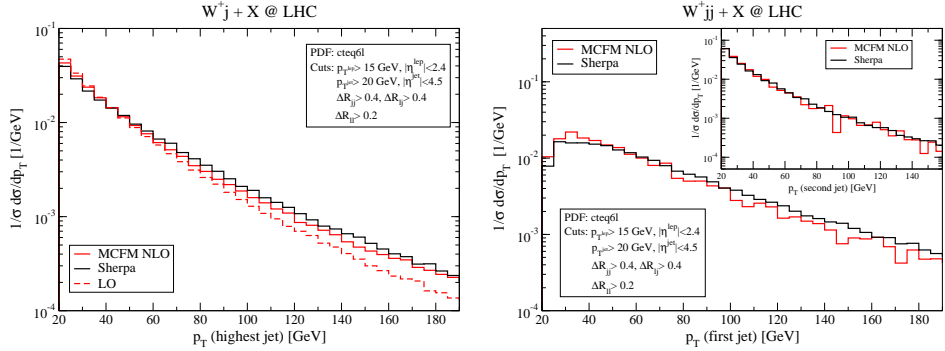

\begin{center}
\begin{tabular}{cc}
\includegraphics[width=6cm]{figures/LHC_Wp1jet_incl.eps}&
\includegraphics[width=6cm]{figures/LHC_Wp2jet_incl.eps}
\end{tabular}
\end{center}
\caption{\label{LHC_inclusive} The $p_T$ distribution of the first jet 
in inclusive $W^+$+1jet (left panel) and the first and second jet in inclusive $W$+2jet 
(right panel) production at the LHC. The \sherpa\ predictions are compared to the 
corresponding NLO QCD prediction obtained with MCFM \cite{Campbell:2002tg}. 
For the naive LO and the NLO calculation the renormalisation and factorisation 
scale has been set to $\mu_R=\mu_F=M_W$.}
\end{figure}

In Fig.\ \ref{LHC_inclusive} inclusive NLO results obtained with MCFM for the $p_\perp$ 
of the hardest jet in $W^+$+1jet and the hardest and second hardest jet in $W^+$+2jets 
events at the LHC are compared to fully inclusive samples generated with \sherpa. There, 
the matrix elements for $W/Z$+0,1,2jet production have been used. The Sudakov and 
$\alpha_s$ reweighted matrix elements have now been combined with the initial and 
final state parton showers. For both cases the \sherpa\ result and the NLO calculation 
are in good agreement. 

Although it should be stressed that the rate predicted by \sherpa\ 
is still a leading order value only, a constant $K$-factor is sufficient to recover 
excellent agreement with a full next-to-leading order calculation for the distributions 
considered. Furthermore, by looking at the inclusive spectra it is obvious that this 
statement still holds true after the inclusion of parton showers and the merging of 
exclusive matrix elements of different jet multiplicities.

\begin{figure}[h]
\begin{center}
\begin{tabular}{cc}
\hspace*{-2mm}\includegraphics[width=6.7cm]{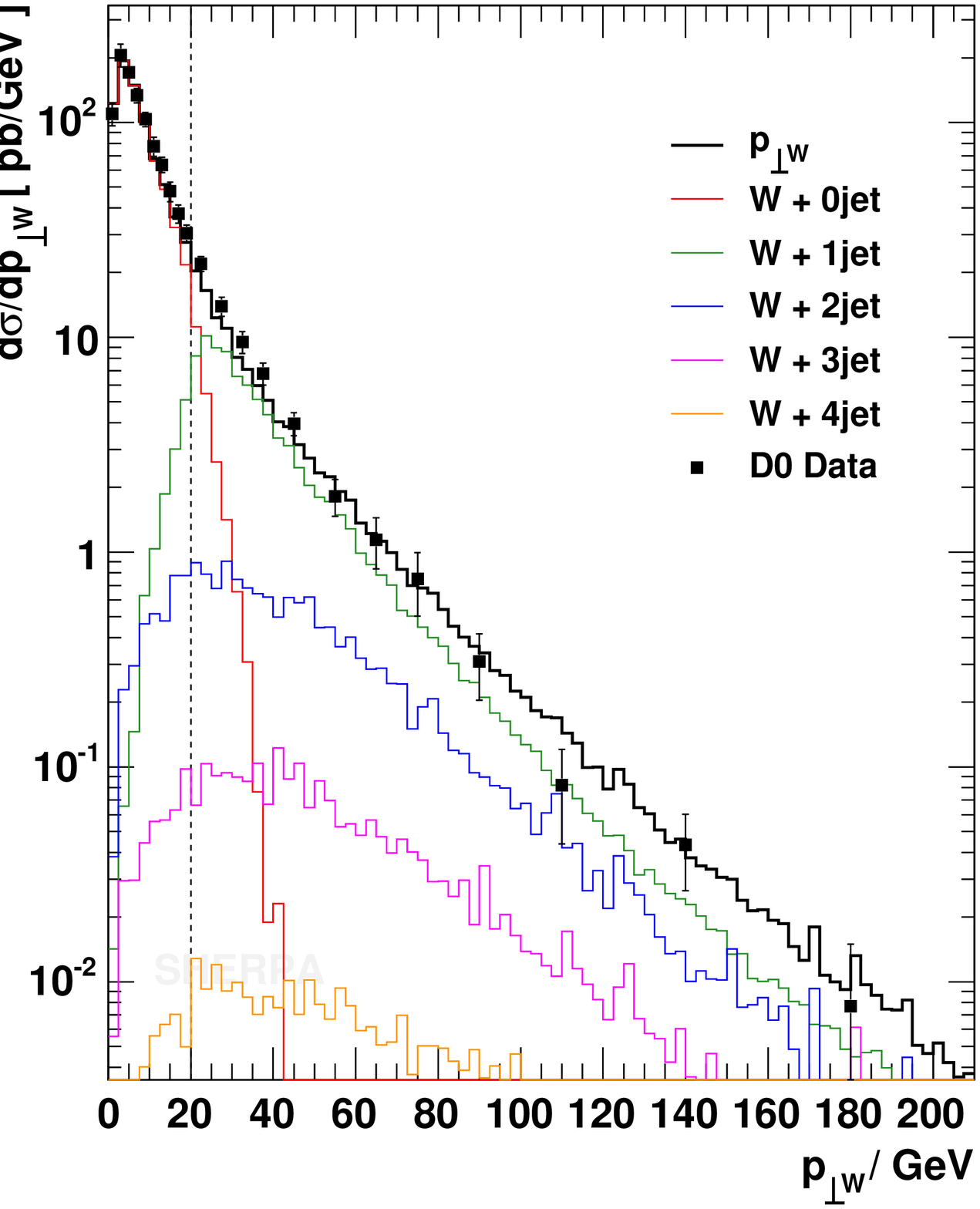}&
\hspace*{-7mm}\includegraphics[width=6.7cm]{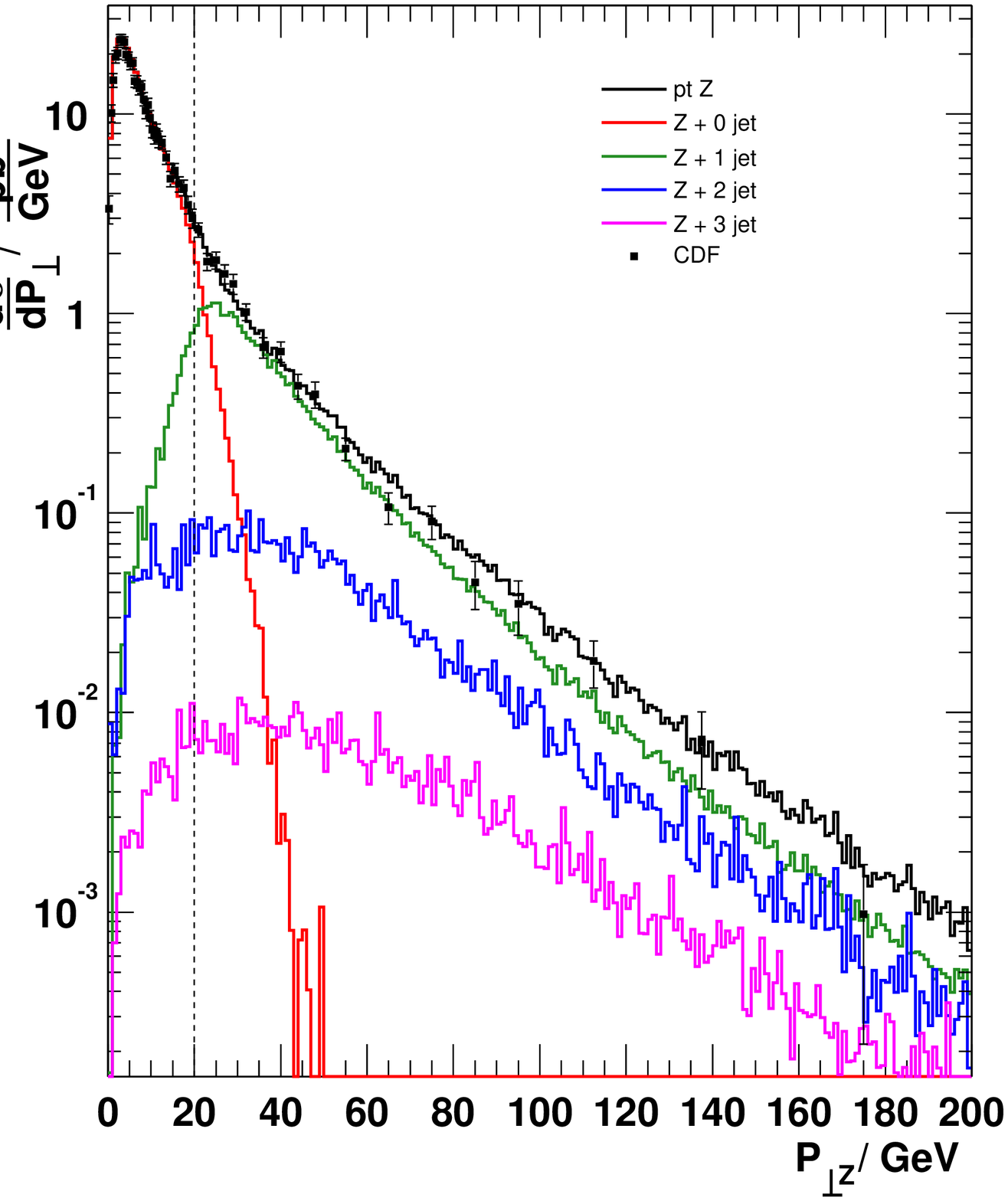}
\end{tabular}
\end{center}
\caption{\label{W_data} The $p_\perp$ distributions of the $W$ (left panel) 
and $Z$ boson (right panel) in comparison with Tevatron Run~I D0 data 
\cite{Abbott:2000xv} and CDF data \cite{Affolder:1999jh}, respectively. The total results 
are indicated by the black lines. The colored lines show the contributions of the different 
multiplicity processes. The applied separation cut for both samples is $Q_{\rm cut}=20$ GeV. }
\end{figure}
Finally, a comparison of the \sherpa\ predictions with experimental data provides 
an ultimate test of \sherpa's ability to describe $W$ and $Z$ production at 
hadron colliders. In Fig.\ \ref{W_data}, the (inclusive) $p_\perp$ distribution 
of the $W$ and $Z$ boson is compared with data from D0 \cite{Abbott:2000xv} 
and CDF \cite{Affolder:1999jh}, respectively, taken at Run~I of the Tevatron. 
After a rescaling of the \sherpa\ predictions by constant $K$-factors the agreement 
with data for both distributions is excellent.

\section{Conclusions}

The results presented for the production of $W$ and $Z$ bosons at hadron colliders prove that the 
merging of tree-level matrix elements and parton showers as implemented in \sherpa\ is 
working in a systematically correct manner; further tests will include, e.g., the sensitivity 
of results to the choice of scale, the quality in describing more complicated correlations, 
for instance of different jets and the simulation of more processes. This agenda currently 
is being worked on, the first results being very encouraging. This indicates that \sherpa\ 
is perfectly suitable to meet the enhanced demands of the community to reliably simulate 
physics processes at the next generation of collider experiments.

\section*{Acknowledgments}
The authors gratefully acknowledge financial support by BMBF, DFG, and GSI. 
S.S.\ wishes to thank the organisers of ``Physics at LHC'' for the extremely 
pleasant atmosphere and fruitful discussions during the conference and the 
financial support provided.
\bigskip
%

%\lastevenpage
\end{document}